\documentclass[aps,preprintnumbers,letterpaper,twocolumn]{revtex4}

\usepackage{epsfig,graphics}
\usepackage{amsfonts,amssymb,amsmath} 
\usepackage{amsthm}         
\usepackage{color}
\newcommand{\chapter}{article}
\newcommand{\comment}[1]{}

\newcommand{\bef}{\rightarrow}
\newcommand{\nbef}{\not\rightarrow}

\theoremstyle{plain}

\theoremstyle{definition}

\begin{document}
\title{A short note on the concept of free choice}
\date{$18^{\text{th}}$ February 2013}
\author{Roger \surname{Colbeck}}
\affiliation{Institute for Theoretical Physics, ETH Zurich, 8093
 Zurich, Switzerland}
\author{Renato \surname{Renner}}
\affiliation{Institute for Theoretical Physics, ETH Zurich, 8093
 Zurich, Switzerland}

\begin{abstract}
  We argue that the concepts of ``freedom of choice'' and of ``causal
  order'' are intrinsically linked: a choice is considered ``free'' if
  it is correlated only to variables in its causal future.
  We discuss the implications of this to Bell-type scenarios, where
  two separate measurements are carried out, neither of which lies in
  the causal future of the other, and where one typically assumes that
  the measurement settings are chosen freely. Furthermore, we refute a
  recent criticism made by Ghirardi and Romano in [arXiv:1301.5040]
  and [arXiv:1302.1635] that we used an unphysical freedom of choice
  assumption in our previous works, [Nat.\ Commun.\
  {\bf 2}, 411 (2011)] and [Phys.\ Rev.\ Lett.\ {\bf 108}, 150402 (2012)].
\end{abstract}

\maketitle

Consider a simple experiment in which a system is prepared in state
$Z$, then a measurement $A$ is chosen and applied to the system, and
finally the outcome $X$ is recorded. How should we express the
requirement that $A$ is a free choice?

We may think of $Z$, $A$, and $X$ as random variables with joint
probability distribution $P_{Z A X}$. For $A$ to be free it is natural
to demand that it can be chosen independently of the state $Z$, i.e.,
$P_{A|Z}=P_A$. However, it would be too restrictive to also require
independence from $X$, i.e., $P_{A|Z X}=P_A$, as we expect the outcome
of an experiment to depend on how we measure. The notion of free
choice is hence intrinsically connected to a causal order: we don't
require that the free choice $A$ is uncorrelated with $X$ since $X$
lies in the causal future of $A$.

\begin{figure}
\includegraphics[clip=true,trim=0cm 0cm 0cm 0cm,width=0.4\textwidth]{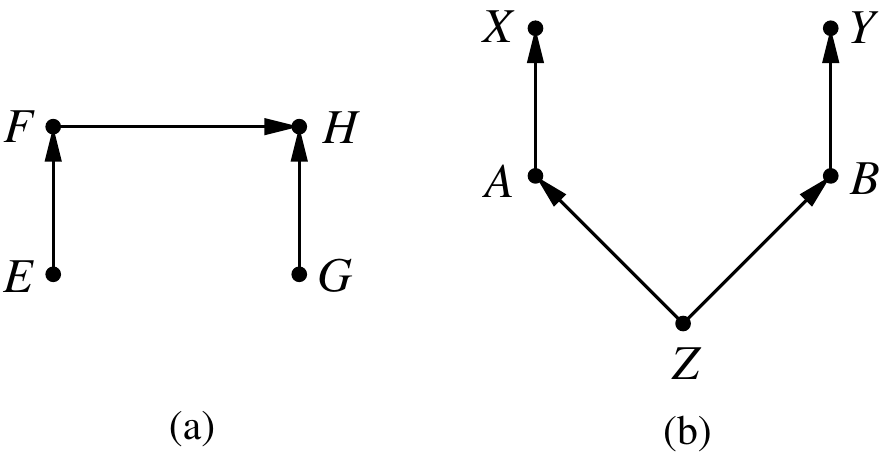}
\caption{Two examples of a causal order.  In (a), $F$ being free
  implies $P_{F|EG}=P_F$, while in (b), if $A$ is free then
  $P_{A|BYZ}=P_A$, for example.}
\label{fig:1}
\end{figure}

This can be easily extended to scenarios involving more than one
measurement. As above, a system's state, measurement choices, and
measurement outcomes may be modeled as random variables, the
collection of which we denote by $\Gamma$. A \emph{causal order} is
then simply a preorder relation~\footnote{A preorder relation is a
  binary relation that is reflexive ($A\bef A$) and transitive (if
  $A\bef B$ and $B\bef C$ then $A\bef C$).} ($\bef$) on $\Gamma$ (see
Fig.~\ref{fig:1} for examples). $A\bef X$ should be interpreted as
``$X$ is in the causal future of $A$''~\footnote{That $X$ is in the
  causal future of $A$ does not mean that $A$ \emph{is} the cause of
  $X$, but only that the causal order does not preclude this.}. While
the causal order may, in principle, be defined arbitrarily, it is
reasonable to demand that it be compatible with \emph{time-ordering},
defined as follows.  For two random variables, $A$ and $X$, the order
$A\bef X$ is taken to hold if and only if $A$ occurs at an earlier
time than the generation of $X$ (with respect to all relativistic
frames~\footnote{Note that relativity theory is not assumed here.
  Relativistic spacetime structure merely offers one possible way to
  physically motivate a causal order.})~\footnote{In other words, $X$
  is \emph{not} in the causal future of $A$ if $X$ occurs at a time
  earlier than $A$ (in some reference frame).}.

Given a set $\Gamma$ with an (arbitrary) causal order, we can define
the concept of a free choice as follows~\footnote{See Definition~4
  of~\cite{CR_new}.}:
\begin{quote}
  A choice $A \in \Gamma$ is \emph{free} if $A$ is uncorrelated with
  the set of all $W \in \Gamma$ that satisfy $A \nbef W$.
\end{quote} 
Said another way, $A$ is free if the only variables it is correlated
with are those it could have caused. Note that the condition $A \nbef
W$ cannot be replaced by $W \bef A$~\footnote{If the condition $A
  \nbef W$ was replaced by $W \bef A$, the requirement for a choice
  $A$ to be free would be that $A$ is uncorrelated only with the set
  of variables in its causal past.  Such an alternative
  characterization would however contradict our intuitive
  understanding of ``freedom of choice'', as can be seen by a simple
  example. Consider the causal order defined by Fig.~\ref{fig:1}(b),
  where $Z$ is the state in which a system has been prepared and where
  $A$ and $B$ are choices made by two different
  experimenters. According to the alternative characterization, $A$
  and $B$ would both be considered free provided $A$ and $B$ are
  independent of $Z$ (i.e., $P_A = P_{A|Z}$ and $P_B = P_{B|Z}$).
  However, this does not prevent $A$ and $B$ from being perfectly
  correlated, a situation in which the two experimenters would likely
  reject any claim that both choices were free.}.

To demonstrate the use of this definition, we consider a Bell-type
setup, where two particles are generated in state $Z$ and subsequently
measured at two distant locations. Let $A$ and $B$ be the choices of
the measurement settings at the two locations, and let $X$ and $Y$ be
the corresponding measurement outcomes.  The two measurements should
be arranged such that neither lies in the causal future of the other,
as depicted in Fig.~\ref{fig:1}(b). We note that, physically, this
causal order can be obtained by carrying out the two measurements in
two spacelike separated regions, and demanding that the causal order
be compatible with time-ordering. Now, assuming that $A$ is free means
that $P_{A|B Y Z} = P_{A}$.

In recent work, we have used the assumption of free choice to show
that there cannot exist any extension of quantum theory with improved
predictive power~\cite{CR_ext}, and that the quantum wave function is
in one-to-one correspondence with its elements of
reality~\cite{CR_wavefn} (see also~\cite{CR_new}). The assumption is
also used in experimental work that provides a fundamental bound on
the maximum probability by which the outcomes of measurements in a
Bell-type setup can be predicted correctly~\cite{SSCRT}. Unhappy with
these consequences, Ghirardi and Romano have, in a sequence of two
papers~\cite{GR1} and~\cite{GR2}, criticized the use of the freedom of
choice assumption in these works, calling it
``unphysical''~\footnote{In addition, in~\cite{GR2}, it is claimed
  that the result of~\cite{CR_wavefn} is based on an extra
  independence assumption of the form $P_{CZ|ABXY}=P_{CZ}$, this
  relation being the authors' interpretation of our remark
  in~\cite{CR_ext} that certain information, $(C,Z)$, is considered
  static. However, this independence is never assumed in our work. The
  implication of our remark is only that information that is used to
  predict the outcome of a measurement should not lie in the causal
  future of the measurement.}. However, the concept of free choice
used in~\cite{CR_ext,CR_wavefn,CR_new,SSCRT} is precisely the one
explained in this note for a causal order compatible with
time-ordering (as defined above), and hence has a clear physical
motivation.

We conclude by remarking that this notion of free choice matches what
Bell said about ``free variables''~\cite{Bell_free}:
\begin{quote}
  For me this means that the values of such variables have
  implications only in their future light cones.
\end{quote}

\noindent{\bf Acknowledgements}|We thank Nicolas Gisin and Sandu
Popescu for discussions on this subject.


\end{document}